\documentclass[reprint,amsmath,amssymb,aps,prl,longbibliography,preprintnumbers,showpacs,floatfix,
nofootinbib]{revtex4-1}
\usepackage{color,graphicx} 
 
\begin{document}

\preprint{DESY 11-029}

\title{ 
 Production of the Exotic $1^{--}$ Hadrons $\phi(2170)$, $X(4260)$ and $Y_b(10890)$
at the LHC and Tevatron via the Drell-Yan Mechanism}

\author{Ahmed~Ali}
\email{ahmed.ali@desy.de}
\author{Wei~Wang}
\email{wei.wang@desy.de}
\affiliation{Deutsches Elektronen-Synchrotron DESY, D-22607 Hamburg, Germany}

\date{\today}

%-------------------------------------------------------
%---------------- ABSTRACT               :--------------
%-------------------------------------------------------
\begin{abstract}
We calculate the Drell-Yan production cross sections and differential distributions in the transverse momentum
and rapidity of the $J^{PC}=1^{--}$ exotic hadrons $\phi(2170)$, $X(4260)$ and $Y_b(10890)$ at the hadron colliders
LHC and the Tevatron. These hadrons are tetraquark (four-quark) candidates, with a hidden $s\bar{s}$, $c\bar{c}$
and $b\bar{b}$ quark pair, respectively. In deriving the distributions and cross sections, we include the
order $\alpha_s$ QCD corrections, resum the large logarithms
in the small transverse momentum region in the impact-parameter formalism, and  use the state 
of the art parton distribution functions.
Taking into account the data on the production and decays of these vector hadrons 
from the $e^+e^-$ experiments, we present the    
 production rates for the processes $pp(\bar{p}) \to
\phi(2170)(\to \phi(1020) \pi^+\pi^- \to K^+K^- \pi^+\pi^-)+...$,  $pp(\bar{p}) \to
X(4260)(\to J/\psi \pi^+\pi^- \to \mu^+\mu^-\pi^+\pi^-)+...$, and
$pp(\bar{p}) \to Y_b(10890)(\to (\Upsilon(1S), \Upsilon(2S), \Upsilon(3S)) \pi^+\pi^- \to \mu^+\mu^-\pi^+\pi^-)+...$.
Their measurements at the hadron colliders will provide new experimental avenues to
explore the underlying dynamics of these hadrons. 
\end{abstract}

%  14.40.Rt Exotic mesons
%13.85.-t Hadron-induced high- and super-high-energy interactions (energy > 10 GeV)
% (for low energies, see 13.75.-n) 
%13.85.N  Inclusive production with identified hadrons

\maketitle

%
%%%%%%%%%%%%%%%%%%%%%%%%%%%%%%%%%%%%%%%%%%%%%%%%%%%%%%%%%%%%%%%%%%%%%%%%%%%%%%%%%
%     Section I : Introduction
%%%%%%%%%%%%%%%%%%%%%%%%%%%%%%%%%%%%%%%%%%%%%%%%%%%%%%%%%%%%%%%%%%%%%%%%%%%%%%%%%
% 
Exotic hadron spectroscopy now stands on firm footing, thanks mainly to experiments during the last
several years at the two $e^+e^-$ B factories, BaBar and Belle, which have reported 
an impressive number of such states in the mass region of the charmonia~\cite{Zupanc:2009qc}.
Most of these states defy a conventional $c\bar{c}$ charmonium interpretation,
 but their {\it affinity} to
decay into the hidden charm states $J/\psi, \psi^\prime$ and into open charm states $D\bar{D}^{(*)}$ reveal that they
have a $c\bar{c}$ component in their Fock space.
Of particular interest for us is the $J^{PC}=1^{--}$ state $Y(4260)$, discovered by BaBar~\cite{Aubert:2005rm}
in the initial state radiation (ISR) process
 $e^+e^- \to \gamma_{\rm ISR} Y(4260) \to \gamma_{\rm ISR} J/\psi \pi^+\pi^-$, 
confirmed later by CLEO~\cite{He:2006kg} and Belle~\cite{:2007sj}, with the latter
finding that  two interfering Breit-Wigner amplitudes to the
$J/\psi \pi^+\pi^-$ state describe the data better.
 Maiani {\it et al.}~\cite{Maiani:2005pe} have interpreted $Y(4260)$
as the first orbital excitation of a diquark-antidiquark (tetraquark) state $([cs][\bar{c}\bar{s}])$.
 Particle Data Group (PDG)~\cite{Nakamura:2010zzi} has assigned the name
$X(4260)$ for this resonance, which is what we also use. 

Evidence exists also for an $s\bar{s}$ state $Y_s(2175)$
with the quantum numbers $J^{PC}=1^{--}$, which was first observed by BaBar~\cite{Aubert:2006bu}
also in the ISR process
$e^+ e^- \to \gamma_{\rm ISR} f_0(980) \phi(1020)$, where $f_0(980)$ is an $0^{++}$ scalar state, 
later confirmed by BESII~\cite{:2007yt} and  Belle~\cite{Shen:2009zze}.
In~\cite{Drenska:2008gr}, $Y_s(2175)$ is interpreted as a tetraquark $[sq][\bar{s}\bar{q}]$
with one unit of relative angular momentum.  
This state is now called $\phi(2170)$ by PDG~\cite{Nakamura:2010zzi}, which we also use.
Likewise, Belle~\cite{Abe:2007tk,:2008pu} measured the state $Y_b(10890)$ in the
process $e^+e^- \to Y_b(10890) \to (\Upsilon(1S), \Upsilon(2S), \Upsilon(3S)) \pi^+\pi^-$.
The production cross sections
and final state distributions, in particular, the dipion invariant mass spectra,
 can be understood if $Y_b(10890)$ is interpreted as a
hidden $\bar bb$ tetraquark state~\cite{Ali:2009es,Ali:2009pi,Ali:2010pq}. 

The aim of this Letter is to investigate the Drell-Yan production of the $J^{PC}=1^{--}$ 
exotic hadrons at the Tevatron and the LHC $p\bar{p}(p) \to \gamma^* \to V +...$,
with $V$ being one of the states $\phi(2170)$, $X(4260)$ or  $Y_b(10890)$.
The running common thread is that all three are candidates for the first orbital excitation of
diquark-antidiquark states with a hidden $s\bar{s}$, $c\bar{c}$ and  $b\bar{b}$ quark content, respectively.
Drell-Yan processes are theoretically better understood than
the corresponding hadronic (prompt) production processes. Unfortunately, due to the very small leptonic branching
 ratios~\cite{Nakamura:2010zzi}, 
production of these exotic states in the traditional $\ell^+\ell^-$ pair ($\ell^\pm=e^\pm,\mu^\pm$) is not promising
in the processes $p\bar{p}(p) \to \gamma^* \to V(\to \ell^+\ell^-) +...$.

 We point out that the corresponding production cross sections
 are large enough to be measured at the LHC and the Tevatron, if, instead of the lepton pair,
one concentrates on the 
 final states in which these exotic vector hadrons have been discovered in the $e^+e^-$ annihilation
experiments: $\phi(2170) \to \phi (1020) f_0(980)$,  $X(4260)\to J/\psi \pi^+\pi^-$, 
and $Y_b(10890) \to (\Upsilon(1S), \Upsilon(2S), \Upsilon(3S))\pi^+\pi^-$.
The obvious advantage is that the essential
input (branching ratios for the discovery channels times the respective leptonic widths) needed for estimating
 the cross sections, are all provided by the $e^+e^-$ experiments, yielding model-independent cross sections  
irrespective of the nature of these states. On the other hand, these measurements
are challenging due to the preponderance of the $\pi^+\pi^-$ pairs from the underlying event in
$pp$ and $p\bar{p}$ collisions, and hence the combinatorial background is expected to be formidable.
 However, we trust that, once the energy-momentum profile of the
background $\pi^+\pi^-$ pairs at the hadron colliders is well understood, the background can be effectively removed by
appropriate cuts, enabling the experiments in carrying out
significant measurements in this sector.

The DY cross sections are based on the factorization theorem (here $X$ denotes a bunch of hadrons)
\begin{eqnarray}
\sigma( pp/p\bar p\to V+X)&=& \int dx_1 dx_2 \sum_{a,b} f_a(x_1) f_b(x_2) \nonumber\\
&&\times  \sigma (a+b\to V (p) +X),
\end{eqnarray}
where $a,b$ denotes a generic parton inside a proton/antiproton, $V=\phi(2170), X(4260), Y_b(10890)$
for the processes considered here with  $p=(p^0,\vec p_T, p^3)$ being the  momentum 4-vector of the
$V$, and $f_a(x_1), f_b(x_2)$ are the parton distribution functions (PDFs), which
depend on the fractional momenta $x_i (i=1,2)$ (an additional scale-dependence is suppressed here). 
We shall adopt the MSTW (Martin-Stirling-Thorne-Watt) PDFs~\cite{Martin:2009iq} in our numerical calculations,
and use another PDF set, the CTEQ10~\cite{Lai:2010vv}, to estimate the uncertainties from this source. 
The process-dependent  partonic cross sections  $\sigma( a +b \to V +X)$ will be computed using the
QCD perturbation theory.

We recall that the leading order contribution  comes from the sub-process $\bar qq\to \gamma^* \to V$ 
\begin{eqnarray}
 \sigma_0 &=& (\delta_{aq}\delta_{b\bar q}+ \delta_{a\bar q}\delta_{b q})\frac{\pi |g_{q\bar qV}|^2}{N_c} \delta(p^2-m_{V}^2),
\end{eqnarray} 
with the color factor $N_c=3$. We include the leading order QCD (i.e., $O(\alpha_s)$) corrections, 
implemented following the pioneering papers ~\cite{Altarelli:1978id,KubarAndre:1978uy}.
%though also $O(\alpha_s^2)$ corrections to the Drell-Yan processes are known.
This formalism is applied to calculate the differential distributions $d^2\sigma/dydp_T^2$,
 with the rapidity defined as $y\equiv \frac{1}{2} \ln \frac{p^0+p^3}{p^0-p^3}$.
The transverse momentum distribution at the tree level has the form $\delta(p_T^2)$. Perturbative QCD
(gluon bremsstrahlung) generates a non-trivial $p_T$-distribution. However, 
 large logarithms of the type $\alpha_s^n\ln^m (p^2/p_T^2)$ arising from higher order QCD corrections 
 spoil the perturbative expansion 
in the small transverse momentum region. These large logarithms  must be resummed in order to improve the convergence
of the perturbation theory. This is done in 
the Collins-Soper-Sterman (CSS) framework~\cite{Collins:1984kg} where the resummation  is carried out in the
 impact parameter space, yielding a simple form for the resummed $p_T$ distribution  
\begin{eqnarray}
 \frac{d^2\sigma}{  dy dp_T^2} &=& 
 \frac{d^2\sigma^{per}}{  dy dp_T^2} + f(p_T)\left(
 \frac{d^2\sigma^{res}}{  dy dp_T^2} -
 \frac{d^2\sigma^{asy} }{  dy dp_T^2}\right)~,\label{eq:resummation}
\end{eqnarray}
in which $ d^2\sigma^{res}/ dy dp_T^2 $ reorganizes the singular terms in the $p_T\to 0$ limit.
Explicitly, this takes the form
\begin{eqnarray} 
 \frac{d^2\sigma^{res}}{  dy dp_T^2}  &= & \frac{\pi^2 }{3s}  \int \frac{ d^2\vec b}{(2\pi)^2} e^{i\vec p_T\cdot  \vec b}
 \sum_{q} g_{q\bar qV}^2\nonumber\\
&\times&  
   \sum_{a,b} \int_{x_1^0}^1 \frac{dx_1}{x_1} f_{a}(x_1,\mu)C_{qa/\bar qa} \left(\frac{x_1^0}{x_1},\mu,g_s,\frac{c}{b}\right)  \nonumber\\
&\times&    \int_{x_2^0}^1 \frac{dx_2}{x_2} f_{b}(x_2,\mu) 
  C_{\bar{q}b/qb} \left(\frac{x_2^0}{x_2}, \mu, g_s,\frac{c}{b}\right)
 \nonumber\\
 &\times& {\rm exp} \Bigg\{-W(b,\frac{c}{b},m_V,x_1,x_2) \Bigg\},\label{eq:explicitresum}
\end{eqnarray}
with $x_1^0 = m_{V}/\sqrt s e^{y}$, $x_2^0 = m_{V}/\sqrt s e^{-y}$, and  $s$ is the square of the center-of-mass
 collision energy. The function $f(p_T)$ in (\ref{eq:resummation})
is introduced as a matching function for which we use~\cite{Kauffman:1991jt}
\begin{eqnarray} 
 f(p_T)= \frac{1}{1+ (p_T/Q_{\rm match})^4}.
\end{eqnarray}
To estimate the uncertainty caused by the matching procedure, we take $Q_{\rm match}=(2\pm 1)m_V$, and this
uncertainty will be
included in the numerical estimates of the transverse momentum distributions.

 The Sudakov factor $W(b, \frac{c}{b},p,x_1,x_2)$ is expressed as 
\begin{eqnarray}
 W(b, \frac{c}{b},p,x_1,x_2)= \int_{\frac{c^2}{b^2}}^{p^2} \frac{d\bar\mu^2}{\bar\mu^2} \Big[ \ln \frac{p^2}{\bar\mu^2} A(g_s(\bar \mu)) +B(g_s(\bar \mu))\Big]~,\nonumber
\end{eqnarray}
and the coefficient functions  $A$, $B$, $C_{qa/\bar{q}a}(x_1^0/x_1)$ and  $C_{qb/\bar{q}b}(x_2^0/x_2)$ are expanded
 (in units of $(\alpha_s/\pi)^{(n)}$). Some
leading terms in these expansions are~\cite{Collins:1984kg}
\begin{eqnarray}
 A^{(1)}= 4/3,\;\;\; B^{(1)}=-2 ,\nonumber\\
C_{jk}^{(0)} =\delta_{jk}\delta(1-z),\;\;\;
C_{jg}^{(1)} = \frac{1}{2} z(1-z),\nonumber\\
C_{jk}^{(1)} =\delta_{jk}\left[\frac{2}{3} (1-z) +\delta(1-z)(\frac{\pi^2}{3}-\frac{8}{3}) \right]~,
\end{eqnarray}
 where the integration constants $C_1,C_2$ in the Sudakov factor (not shown explicitly)
 and the renormalization scale $\mu$ in (\ref{eq:explicitresum}) have been taken as $C_1=\mu b=c=2e^{-\gamma_E}$ and $C_2=1$,
where $\gamma_E=0.57722$ is the Euler constant.  
 
The asymptotic term in (\ref{eq:resummation}) coincides with the perturbative results in the small $p_T$ region
\begin{eqnarray}
 \frac{d^2\sigma^{asy} }{  dy dp_T^2}=\left.
 \frac{d^2\sigma^{per}}{  dy dp_T^2}\right|_{p_T^2\to 0}~,
\end{eqnarray}
so that in this region the resummed terms dominate.  The factorization scale is chosen as
 $\mu=\sqrt {p_T^2+m_V^2}$. 

As the large impact parameter $b$ corresponds to a low momentum scale, 
a cutoff  is introduced in the CSS formalism~\cite{Collins:1984kg}, which replaces the parameter $b$  by
 $b^*= b/\sqrt {1+b^2/b_{max}^2}$, with  $b_*$ bounded from above by  $b_{max}$.
The non-perturbative effects to compensate this cutoff are incorporated into a phenomenological function
 $F_{NP}(b,m_V,x_1,x_2)$, and a
 commonly-adopted parametrization  obtained by fitting the data on $W$  and $Z$
 production~\cite{Ladinsky:1993zn} is given by
\begin{eqnarray}
 F_{NP}&=& {\rm exp}\left[- g_1b^2 -g_2b^2 \ln \frac{m_V}{2Q_0}-g_1g_3 b\ln (\frac{x_1x_2}{0.01})\right]~,\nonumber
%% from Eq.(2.2)  of \cite{Ladinsky:1993zn} 
\end{eqnarray}
where $g_1=0.11~{\rm GeV}^2$, $g_2=0.58~{\rm GeV}^2$, $g_3=-1.5~{\rm GeV}^{-1}$  and $Q_0=1.6$ GeV for
 $b_{max}=0.5 {\rm GeV}^{-1}$. It should be pointed out that the above
value of $Q_0$ is not appropriate for $\phi(2170)$, as in this case $m_{\phi(2170)}<2Q_0$, which 
would lead to an enhancement of the
 large $b$-region instead of  suppressing it, and therefore in our calculation we use as input
 $Q_0=1.0$ GeV, which we adopt for the $Y_b(10890)$ and $X(4260)$ cases as well.  
  
The electromagnetic coupling constants $g_{q \bar q V}$
 are related to the $e^+e^- V$ coupling $g_{ee V}$ by $g_{q \bar q V}= e_qg_{ee V}$.
The relevant $e^+e^-$  experimental data which are used to derive these parameters  are collected in
 Table~\ref{tab:inputs}. The entries for $\Gamma_{ee}(Y_b) {\cal B}(Y_b \to \Upsilon(nS) \pi^+\pi^-)$
 are obtained by using the
relation $\Gamma_{ee}(Y_b)   {\cal B}(Y_b \to \Upsilon(nS) \pi^+\pi^- )=\Gamma_{Y_b}m_{Y_b}^2\sigma(\Upsilon(nS) \pi^+\pi^-)
/(12\pi)$, with all three quantities
on the r.h.s. taken from Belle~\cite{:2008pu}.

%
%%%%%%%%%%%%%%%%%%%%%%%%%%%%%%%%%%%%%%%%%%%%%%%%%%%%%%%%%%%%%%%%%%%%%%%%%%%%%%%%%
%     Section III: Numerical Estimates
%%%%%%%%%%%%%%%%%%%%%%%%%%%%%%%%%%%%%%%%%%%%%%%%%%%%%%%%%%%%%%%%%%%%%%%%%%%%%%%%%
%

\begin{table}
\caption{Masses, total  and partial decay widths of the $\phi(2170)$, $X(4260)$ and $Y_b(10890)$.
 Unless specified, all input values are taken from the PDG review~\cite{Nakamura:2010zzi}  }\label{tab:inputs}
\begin{tabular}{|cccc|}
\hline\hline
 &$m_V$ (MeV) &$\Gamma$  (MeV) & $\Gamma_{ee} {\cal B}$ (eV)\\\hline
 $\phi(2170)$   &$2175\pm 15$&$61\pm 18$ &
$2.5\pm 0.9$~\footnote{$\Gamma_{ee}\times {\cal B}(\phi(2170)\to \phi(1020) f_0(980))$.}  \\   
$ X(4260)$  & $4263^{+8}_{-9}$  & $108\pm 21$~\cite{:2007sj} &$6.0^{+4.9}_{-1.3}$~\footnote{$\Gamma_{ee}\times {\cal B}(X(4260)\to J/\psi  \pi^+\pi^-)$, corresponding to Solution I.} ~\cite{:2007sj}\\
 $Y_{b}(10890)$ & $10888.4^{+3.0}_{-2.9}$~\cite{:2008pu} & $30.7^{+8.9}_{-7.7}$~\cite{:2008pu}&
 $0.69^{+0.23}_{-0.20}$~\footnote{$\Gamma_{ee}\times {\cal B}(Y_b(10890)\to \Upsilon(1S)
  \pi^+\pi^-)$ obtained from
  $\sigma=(2.78^{+0.48}_{-0.41})$ pb. For $Y_b\to \Upsilon(2S)\pi^+\pi^-$, the cross section $(4.82^{+1.01}_{-0.91})$ pb gives $\Gamma_{ee}  {\cal B}=(1.20^{+0.43}_{-0.37})$ eV,  while for $Y_b\to \Upsilon(3S)\pi^+\pi^-$, the cross section $(1.71^{+0.42}_{-0.39})$ pb corresponds to $\Gamma_{ee}  {\cal B}=(0.42^{+0.16}_{-0.14})$ eV. }~\cite{:2008pu}\\\hline
${\cal B}_{\phi \to K^+K^-}$ &$ (48.9\pm0.5)\%$&
${\cal B}_{f_0(980)\to \pi^+\pi^-}$ &$ (50^{+7}_{-9})\%$~\cite{Ablikim:2004cg}\\
${\cal B}_{J/\psi \to \mu^+\mu^-}$ &$ (5.93\pm0.06)\%$&
${\cal B}_{\Upsilon(1S)\to \mu^+\mu^-}$ &$ (2.48\pm0.05)\%$\\ 
${\cal B}_{\Upsilon(2S)\to \mu^+\mu^-}$ &$ (1.93\pm0.17)\%$ & \ 
${\cal B}_{\Upsilon(3S)\to \mu^+\mu^-}$ &$ (2.18\pm0.21)\%$
\\\hline  
\hline 
\end{tabular} 
\end{table}

\begin{figure*}[!t]
\includegraphics[width=0.25\textwidth]{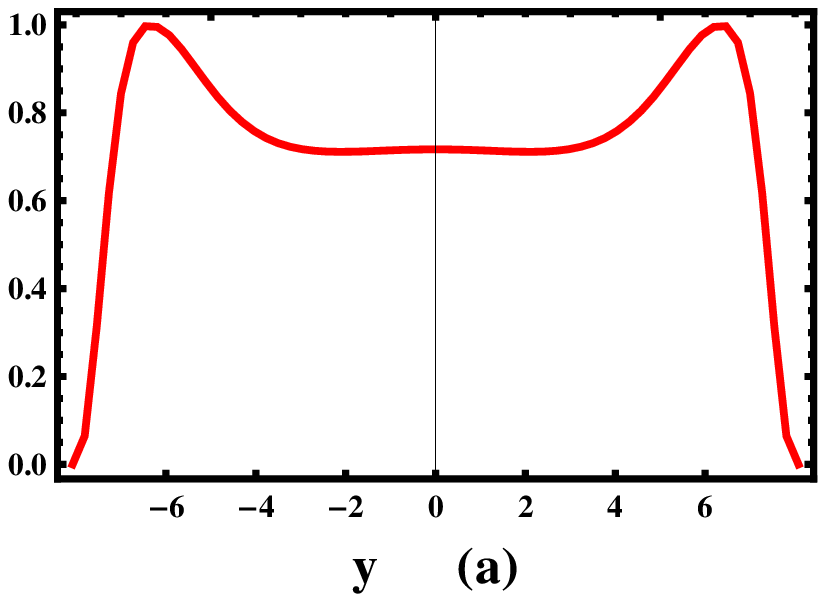}
\includegraphics[width=0.25\textwidth]{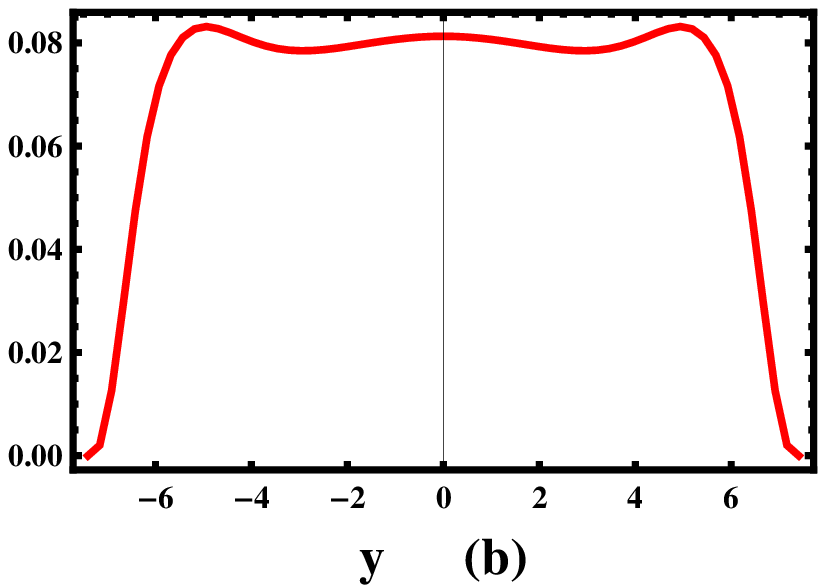}\hspace{2mm}
\includegraphics[width=0.26\textwidth]{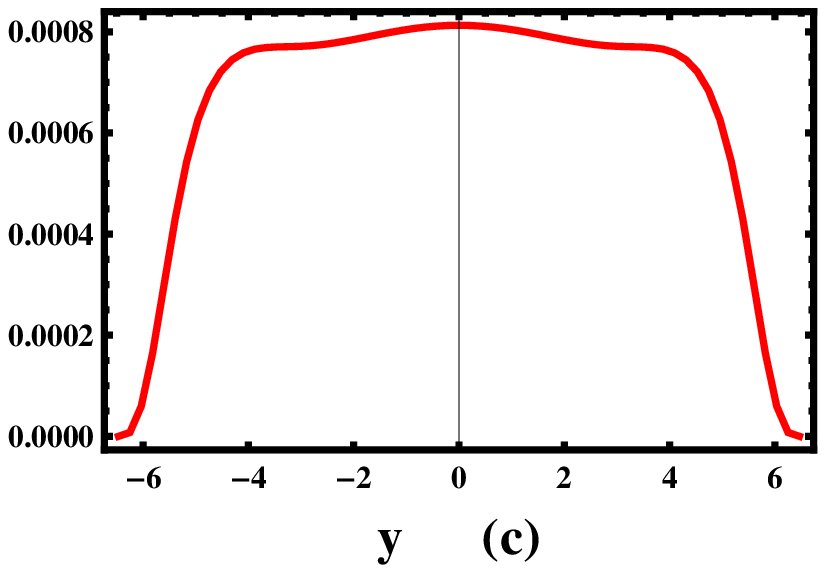}\hspace{2mm}
\caption{Rapidity distributions  $\frac{d\sigma}{dy} $ (in units of pb) for the process 
 (a) $pp\to( \phi(2170) \to  \phi(1020) f_0(980)\to K^+K^- \pi^+\pi^-) + ...$,
 (b) $pp\to( X(4260)\to  J/\psi \pi^+\pi^- \to \mu^+\mu^- \pi^+\pi^-) + ...$,
 and (c) $pp\to( Y_b(10890) \to  \Upsilon (1S,2S,3S)\pi^+\pi^- \to \mu^+\mu^- \pi^+\pi^-) + ...$
  at LHC with $\sqrt s=$ 7TeV  using the MSTW PDFs. }
\label{fig:rapidity}
\end{figure*}

\begin{table}
\caption{Cross sections  (in units of pb)  for the processes
$p \bar{p}(p) \to \phi(2170) (\to  \phi(1020) f_0(980) \to K^+K^- \pi^+\pi^-)$,
  $ p \bar{p}(p) \to X(4260)(\to  J/\psi \pi^+\pi^- \to \mu^+\mu^- \pi^+\pi^-)$, and
 $ p \bar{p}(p) \to  Y_b(10890) (\to  \Upsilon (1S,2S,3S)\pi^+\pi^- \to \mu^+\mu^- \pi^+\pi^-)$,  
  at the Tevatron ($\sqrt s=$ 1.96 TeV)
 and the LHC ($\sqrt s=$ 7 TeV and 14 TeV),  using the MSTW PDFs.  A rapidity range ($|y| < 2.5 $) is assumed for the
Tevatron experiments (CDF and D0) and for the LHC experiments (ATLAS and CMS);  
 a rapidity range $1.9<y<4.9$ is used for the LHCb. }\label{tab:crosssection}
\begin{tabular}{|c|c|c|c|}
\hline\hline
&$\phi(2170)$ &$ X(4260)$ 
&$Y_{b}(10890)$  
\\\hline
 Tevatron$(|y|<2.5)$  &$2.3^{+0.9}_{-0.9}$ &$0.23^{+0.19}_{-0.05}$    &$0.0020^{+0.0006}_{-0.0005} $
\\
LHC 7TeV $(|y|<2.5)$  &$3.6^{+1.4}_{-1.4}$ &$0.40^{+0.32}_{-0.09}$  &$0.0040^{+0.0013}_{-0.0011}$ 
\\ 
LHCb 7TeV   ($1.9<y<4.9$) &$2.2^{+1.2}_{-1.1}$&$0.24^{+0.20}_{-0.07}$ &$0.0023^{+0.0007}_{-0.0006}$ 
\\ 
LHC 14TeV $(|y|<2.5)$  &$4.5^{+1.9}_{-1.9}$ &$0.54^{+0.44}_{-0.12}$  &$0.0060^{+0.0019}_{-0.0016}$ 
\\ 
LHCb  14TeV  ($1.9<y<4.9$) &$2.7^{+1.9}_{-1.6}$&$0.31^{+0.27}_{-0.11}$ &$0.0033^{+0.0011}_{-0.0010}$ 
\\
\hline\hline
\end{tabular} 
\end{table}

\begin{figure*}[!t]
\includegraphics[width=0.25\textwidth]{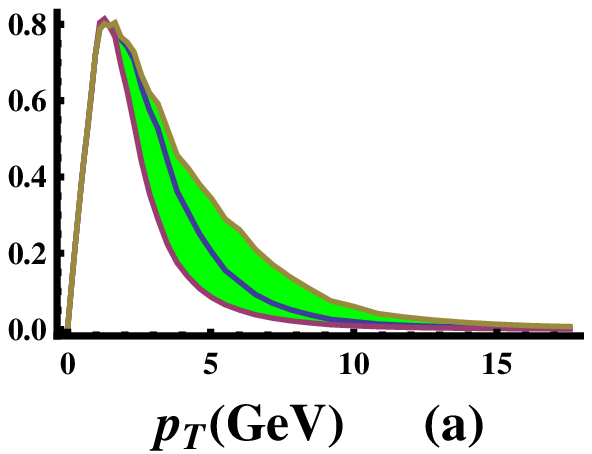}\hspace{2mm} 
\includegraphics[width=0.25\textwidth]{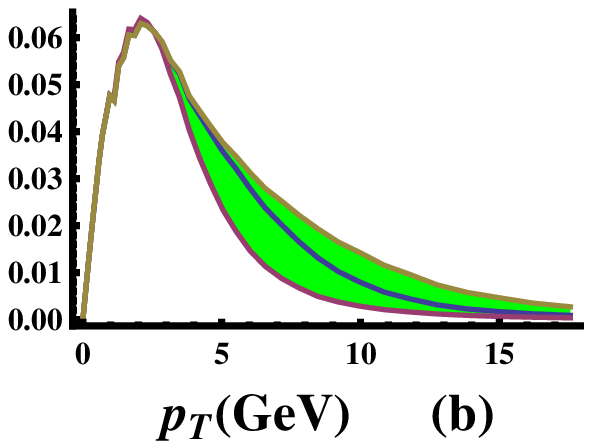}\hspace{2mm}
\includegraphics[width=0.25\textwidth]{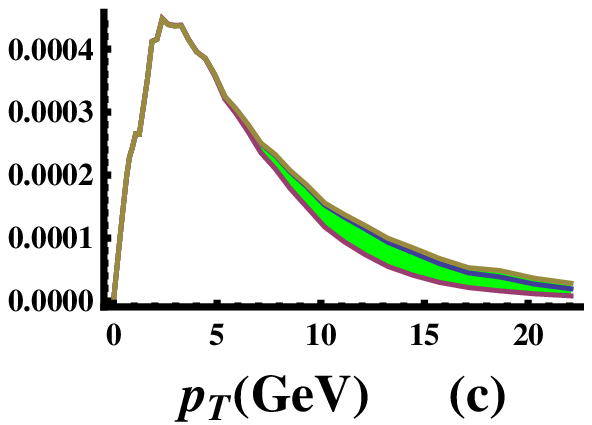}
\caption{Transverse momentum  distributions $\frac{d\sigma}{dp_T}$ (in units of ${\rm pb/GeV}$) for the process 
 (a) $pp\to( \phi(2170) \to  \phi(1020) f_0(980) \to K^+K^- \pi^+\pi^-)+ ... $, 
 (b) $pp\to( X(4260)\to  J/\psi \pi^+\pi^- \to \mu^+\mu^- \pi^+\pi^-) + ... $,
 and (c) $pp\to( Y_b(10890) \to  \Upsilon (1S,2S,3S)\pi^+\pi^- \to \mu^+\mu^- \pi^+\pi^-) + ... $ 
 at the LHC ( $\sqrt s=$ 7 TeV)
with the rapidity cut  $|y|<2.5$ using the MSTW PDFs.
 Uncertainties caused by the matching functions are displayed through $1/[1+(p_T/Q_{\rm match})^4]$ with
 $Q_{\rm match}=(2\pm 1) m_{V}$. }
\label{fig:transverse-momentum}
\end{figure*}

Having specified the formalism and the necessary inputs, we present our numerical results. As the distributions
at the Tevatron and the LHC are rather similar, we show the figures only for the LHC.
Rapidity distributions $d\sigma/dy$ (in units of pb) for the  three Drell-Yan processes at the LHC 
for $\sqrt{s}=7$ TeV are shown in  
Fig.~\ref{fig:rapidity}:
 (a) $pp\to( \phi(2170) \to  \phi(1020) f_0(980) \to K^+K^- \pi^+\pi^-)+ ... $, 
 (b) $pp\to( X(4260)\to  J/\psi \pi^+\pi^- \to \mu^+\mu^- \pi^+\pi^-)+ ... $,
 and (c) $pp\to( Y_b(10890) \to  \Upsilon (1S,2S,3S)\pi^+\pi^- \to \mu^+\mu^- \pi^+\pi^-)+ ... $
(contributions from three intermediate states have been added).
 The normalized distributions are stable, though the indicated uncertainties
in the normalization in Table~\ref{tab:crosssection} discussed below will also reflect in the rapidity distributions
shown in this figure. The corresponding transverse momentum distributions $d\sigma/dp_T$ (in units of
pb/GeV) are shown in Fig.~\ref{fig:transverse-momentum}, which are
obtained for the rapidity range $|y|< 2.5$ (for ATLAS and CMS). The corresponding distributions in the
 rapidity range $1.9 < y < 4.9$ (for the LHCb) are very similar, and hence not shown.
 The uncertainties caused by the matching
 functions are displayed.

The integrated cross sections for the processes
 $ pp(\bar{p}) \to  Y_b(10890) (\to  \Upsilon (1S,2S,3S)\pi^+\pi^- \to \mu^+\mu^- \pi^+\pi^-)+...$,
   $ pp(\bar{p}) \to X(4260)(\to  J/\psi \pi^+\pi^- \to \mu^+\mu^- \pi^+\pi^-)+...$, and 
 $ pp(\bar{p}) \to \phi(2170) (\to  \phi(1020) f_0(980) \to K^+K^- \pi^+\pi^-)+...$  at the Tevatron ($\sqrt s=$ 1.96 TeV) and
the LHC (for $\sqrt s=$ 7 TeV and $14$ TeV) are presented  in Table~\ref{tab:crosssection}, using the MSTW PDFs~\cite{Martin:2009iq}.
 The errors shown are from the parametric uncertainties in the PDFs and the
various experimental inputs given in Table \ref{tab:inputs}, which we have added in quadrature.
We have also checked that our results are modified only moderately if we  use a different set of PDFs.
 For the CTEQ10 PDFs~\cite{Lai:2010vv}, most  changes amount to less than $30\%$, which are
 smaller than the uncertainties from the experimental input.
 We remark that the cross sections for CDF and D0 ($\sqrt{s}=1.96$ TeV) and the LHCb  
(for $\sqrt{s}=7$ TeV) are comparable, despite different center-of-mass energies,  due to their 
different rapidity ranges, whereas the cross sections
 for the ATLAS and CMS detectors at the LHC 
 are larger by typically
1.6 (for $\phi(2170)$), 1.7 (for $ X(4260)$) and 2.0 (for $Y_b(10890)$), compared to the ones
calculated for the CDF and D0 at the Tevatron.  Another remark concerns the collision energy dependence. The cross sections at the LHC  with $\sqrt s=14$ TeV are  enhanced by roughly
1.2 (for $\phi(2170)$), 1.3 (for $ X(4260)$) and 1.5 (for $Y_b(10890)$) compared to
 the corresponding results at $\sqrt s=$ 7TeV. 

 To estimate the number of events, we
assume an integrated luminosity of 10 fb$^{-1}$ at the Tevatron by the end of this year, and half that number at
the LHC  (for $\sqrt s=$ 7 TeV) by the end of 2012. This yields $2.3 \times 10^4$ events for the mode
 $\phi(2170) \to \phi(1020) f_0(980) \to K^+ K^- \pi^+\pi^-$,  $2.3 \times 10^3$ events for the mode 
 $X(4260) \to J/\psi \pi^+\pi^-\to \mu^+\mu^-$ (and approximately the same number for the  
 $X(4260) \to J/\psi \pi^+\pi^-\to e^+ e^-\pi^+\pi^-$ mode), and only about 20 events for the mode
$Y_b(10890) \to (\Upsilon(1S), \Upsilon(2S),\Upsilon(3S))\pi^+\pi^- \to \mu^+ \mu^- \pi^+\pi^-$ (and
approximately the same number of events for the $Y_b(10890) \to (\Upsilon(1S), \Upsilon(2S),\Upsilon(3S))\pi^+\pi^-
 \to e^+ e^- \pi^+\pi^-$ mode). The corresponding numbers for the ATLAS and CMS [LHCb] are $1.8 [1.1]\times 10^4$,
  $2.0 [1.2] \times 10^{3}$, and 20 [11], respectively. Hence, all these processes have measurable rates, given
the luminosities at the Tevatron and the LHC, though the measurement of $Y_b(10890)$ in the Drell-Yan process
may have to wait for higher luminosities and/or higher center-of-mass energy at the LHC. 
 
In summary, we have presented the Drell-Yan cross sections and the corresponding differential distributions
 for the production of
the $J^{PC}=1^{--}$ exotic vector hadrons $\phi(2170)$, $X(4260)$ and $Y_b(10890)$ at the Tevatron and the LHC.
 The estimates given here are model-independent due to the
 experimental
input provided by the $e^+e^-$ experiments. To unravel the dynamics underlying the exotic spectroscopy, one
will have to undertake  detailed dynamical studies involving the final states.
% spectra~\cite{Ali:2009es,Ali:2010pq}.

We acknowledge helpful discussions with Silja Brensing, Christian Hambrock and Satoshi Mishima. W. W. is supported by
 the Alexander-von-Humboldt Stiftung.


\begin{thebibliography}{99}

%\cite{Zupanc:2009qc}
\bibitem{Zupanc:2009qc}
  For a recent experimental review, see A.~Zupanc  [Belle Collaboration],
  %``Hadron Spectroscopy Results from Belle,''
  arXiv:0910.3404 [hep-ex].
  %%CITATION = ARXIV:0910.3404;%%

%%% Y(4260)) BaBar discovery 
%\cite{Aubert:2005rm}
\bibitem{Aubert:2005rm}
  B.~Aubert {\it et al.}  [BABAR Collaboration],
  %``Observation of a broad structure in the pi+ pi- J / psi mass spectrum
  %around 4.26-Gev/c**2,''
  Phys.\ Rev.\ Lett.\  {\bf 95}, 142001 (2005).
 % [arXiv:hep-ex/0506081].
  %%CITATION = PRLTA,95,142001;%%

%\cite{He:2006kg}
\bibitem{He:2006kg}
  Q.~He {\it et al.}  [CLEO Collaboration],
  %``Confirmation of the Y(4260) resonance production in ISR,''
  Phys.\ Rev.\  D {\bf 74}, 091104 (2006).
% [arXiv:hep-ex/0611021].
  %%CITATION = PHRVA,D74,091104;%%

%%%%% belle Yc(4008) and Yc(4260) 
%\cite{:2007sj}
\bibitem{:2007sj}
  C.~Z.~Yuan {\it et al.}  [Belle Collaboration],
  %``Measurement of $e^+e^- \to \pi^+\pi^-J/\psi$ Cross Section via Initial
  %State Radiation at Belle,''
  Phys.\ Rev.\ Lett.\  {\bf 99}, 182004 (2007).
%  [arXiv:0707.2541 [hep-ex]].
  %%CITATION = PRLTA,99,182004;%%

%%%%Tetraquark 4260
%\cite{Maiani:2005pe}
\bibitem{Maiani:2005pe}
  L.~Maiani, V.~Riquer, F.~Piccinini and A.~D.~Polosa,
  %``Four Quark Interpretation of Y(4260),''
  Phys.\ Rev.\  D {\bf 72}, 031502 (2005).
%  [arXiv:hep-ph/0507062].
  %%CITATION = PHRVA,D72,031502;%%

%%%% PDG
%\cite{Nakamura:2010zzi}
\bibitem{Nakamura:2010zzi}
  K.~Nakamura {\it et al.}  [Particle Data Group],
  %``Review of particle physics,''
  J.\ Phys.\ G {\bf 37}, 075021 (2010).
  %%CITATION = JPHGB,G37,075021;%%


%\cite{Aubert:2006bu}
\bibitem{Aubert:2006bu}
  B.~Aubert {\it et al.}  [BABAR Collaboration],
  %``A Structure at 2175-MeV in $e^{+} e^{-} \to \phi$ f0(980) Observed via
  %Initial-State Radiation,''
  Phys.\ Rev.\  D {\bf 74}, 091103 (2006).
%  [arXiv:hep-ex/0610018].
  %%CITATION = PHRVA,D74,091103;%%

%\cite{:2007yt}
\bibitem{:2007yt}
  M.~Ablikim {\it et al.}  [BES Collaboration],
  %``Observation of $Y(2175)$ in $J/\psi\to \eta\phi f_0(980)$,''
  Phys.\ Rev.\ Lett.\  {\bf 100}, 102003 (2008).
%  [arXiv:0712.1143 [hep-ex]].

%\cite{Shen:2009zze}
\bibitem{Shen:2009zze}
  C.~P.~Shen {\it et al.}  [Belle Collaboration],
  %``Observation of the $\phi(1680)$ and the Y(2175) in $e^+ e^- \to
  %\phi\pi^+\pi^-$,''
  Phys.\ Rev.\  D {\bf 80}, 031101 (2009).
%  [arXiv:0808.0006 [hep-ex]].
  %%CITATION = PHRVA,D80,031101;%%

%\cite{Drenska:2008gr}
\bibitem{Drenska:2008gr}
  N.~V.~Drenska, R.~Faccini, A.~D.~Polosa,
  %``Higher Tetraquark Particles,''
  Phys.\ Lett.\  {\bf B669}, 160-166 (2008).
%  [arXiv:0807.0593 [hep-ph]].


%\cite{Abe:2007tk}
\bibitem{Abe:2007tk}
  K.~F.~Chen {\it et al.}  [Belle Collaboration],
  %``Observation of anomalous $\Upsilon(1S) \pi^+ \pi^-$ and
  %$\Upsilon(2S) \pi^+ \pi^-$  production near the $\Upsilon(5S)$
  %resonance,'' 
  Phys.\ Rev.\ Lett.\  {\bf 100}, 112001 (2008).
 % [arXiv:0710.2577 [hep-ex]].
  %%CITATION = PRLTA,100,112001;%%

%\cite{:2008pu}
\bibitem{:2008pu}
  I.~Adachi {\it et al.}  [Belle Collaboration],
  %``Observation of an enhancement in e+e- to Upsilon(1S)pi+pi-,
  %Upsilon(2S)pi+pi-, and Upsilon(3S)pi+pi- production at Belle,''
%  arXiv:0808.2445 [hep-ex];
  Phys.\ Rev.\  D {\bf 82}, 091106 (2010).
  %%CITATION = ARXIV:0808.2445;%%


%\cite{Ali:2009es}
\bibitem{Ali:2009es}
  A.~Ali, C.~Hambrock, and M.~J.~Aslam,
  %``Tetraquark interpretation of the BELLE data on the anomalous
  %$\Upsilon(1S) \pi^+\pi^-$ and $\Upsilon(2S) \pi^+\pi^-$ production
  %near the $\Upsilon(5S)$ resonance,''
  Phys.\ Rev.\ Lett.\  {\bf 104}, 162001 (2010).
  %[arXiv:0912.5016 [hep-ph]].
  %%CITATION = PRLTA,104,162001;%%

%\cite{Ali:2009pi}
\bibitem{Ali:2009pi}
  A.~Ali, C.~Hambrock, I.~Ahmed, and M.~J.~Aslam,
  %``A case for hidden $b\bar{b}$ tetraquarks based on $e^+e^- \to
  %b\bar{b}$ cross section between $\sqrt{s}=10.54$ and 11.20 GeV,''
  Phys.\ Lett.\  B {\bf 684}, 28 (2010).
  %[arXiv:0911.2787 [hep-ph]].
  %%CITATION = ARXIV:0911.2787;%%

%\cite{Ali:2010pq}
\bibitem{Ali:2010pq}
  A.~Ali, C.~Hambrock and S.~Mishima,
  %``Tetraquark-based analysis and predictions of the cross sections and
  %distributions for the processes e^+ e^- --> Upsilon(1S) (pi^+ pi^-, K^+ K^-,
  %eta pi^0) near Upsilon(5S),''
  Phys.\ Rev.\ Lett.\  {\bf 106}, 092002 (2011).
 %  [arXiv:1011.4856 [hep-ph]].
  %%CITATION = PRLTA,106,092002;%%

%% MSTW PDFs
%\cite{Martin:2009iq}
\bibitem{Martin:2009iq}
  A.~D.~Martin, W.~J.~Stirling, R.~S.~Thorne and G.~Watt,
  %``Parton distributions for the LHC,''
  Eur.\ Phys.\ J.\  C {\bf 63}, 189 (2009).
 % [arXiv:0901.0002 [hep-ph]].
  %%CITATION = EPHJA,C63,189;%%

%%%% CTEQ 10 PDFs
%\cite{Lai:2010vv}
\bibitem{Lai:2010vv}
  H.~L.~Lai {\it et. al},
% M.~Guzzi, J.~Huston, Z.~Li, P.~M.~Nadolsky, J.~Pumplin and C.~P.~Yuan,
  %``New parton distributions for collider physics,''
  Phys.\ Rev.\  D {\bf 82}, 074024 (2010).
%  [arXiv:1007.2241 [hep-ph]].
  %%CITATION = PHRVA,D82,074024;%%

%%% Drell-Yan NLO
%\cite{Altarelli:1978id}
\bibitem{Altarelli:1978id}
  G.~Altarelli, R.~K.~Ellis and G.~Martinelli,
  %``Leptoproduction And Drell-Yan Processes Beyond The Leading Approximation In
  %Chromodynamics,''
  Nucl.\ Phys.\  B {\bf 143}, 521 (1978)
  [Erratum-ibid.\  B {\bf 146}, 544 (1978)];
  Nucl.\ Phys.\  B {\bf 157}, 461 (1979).
  %%CITATION = NUPHA,B143,521;%%

%\cite{KubarAndre:1978uy}
\bibitem{KubarAndre:1978uy}
  J.~Kubar-Andre and F.~E.~Paige,
  %``Gluon Corrections To The Drell-Yan Model,''
  Phys.\ Rev.\  D {\bf 19}, 221 (1979).
  %%CITATION = PHRVA,D19,221;%%


%%%% CSS resummation 
%\cite{Collins:1984kg}
\bibitem{Collins:1984kg}
  J.~C.~Collins, D.~E.~Soper and G.~F.~Sterman,
  %``Transverse Momentum Distribution In Drell-Yan Pair And W And Z Boson
  %Production,''
  Nucl.\ Phys.\  B {\bf 250}, 199 (1985).
  %%CITATION = NUPHA,B250,199;%%



%%%%%%%%%%matching function Eq.(4.5)
%\cite{Kauffman:1991jt}
\bibitem{Kauffman:1991jt}
  R.~P.~Kauffman,
  %``Higgs boson p(T) in gluon fusion,''
  Phys.\ Rev.\  D {\bf 44}, 1415 (1991).
  %%CITATION = PHRVA,D44,1415;%%



%%%%% FNP parametrization 
%\cite{Ladinsky:1993zn}
\bibitem{Ladinsky:1993zn}
  G.~A.~Ladinsky and C.~P.~Yuan,
  %``The Nonperturbative regime in QCD resummation for gauge boson production at
  %hadron colliders,''
  Phys.\ Rev.\  D {\bf 50}, 4239 (1994).
 % [arXiv:hep-ph/9311341].
  %%CITATION = PHRVA,D50,4239;%%


%\cite{Ablikim:2004cg}
\bibitem{Ablikim:2004cg}
  M.~Ablikim {\it et al.}  [BES Collaboration],
  %``Evidence for f0(980)f0(980) production in chi_c0 decays,''
  Phys.\ Rev.\  D {\bf 70}, 092002 (2004);
%  [arXiv:hep-ex/0406079];
%  %%CITATION = PHRVA,D70,092002;%%
  %``Partial Wave Analysis of $\chi_{c0}\to\pi^+\pi^-K^+K^-$,''
  Phys.\ Rev.\  D {\bf 72}, 092002 (2005).
%  [arXiv:hep-ex/0508050].
  %%CITATION = PHRVA,D72,092002;%%


\end{thebibliography}
\end{document}